\documentclass[prb,twocolumn,showpacs,superscriptaddress]{revtex4-2}

\usepackage{graphicx}
\usepackage{setspace}
\usepackage{amsmath}
\usepackage{mathtools}
\usepackage{tabularx}
\usepackage{mathrsfs}
\usepackage{lipsum}
\usepackage{xcolor}
\usepackage[citecolor=blue,linkcolor=blue,colorlinks=true,linkbordercolor=white]{hyperref}
\usepackage{float}

\newcommand{\be}{\begin{equation}}
\newcommand{\ee}{\end{equation}}
\newcommand{\bea}{\begin{eqnarray}}
\newcommand{\eea}{\end{eqnarray}}

\newcommand{\la}{\langle}
\newcommand{\ra}{\rangle}
\newcommand{\ua}{\uparrow}
\newcommand{\da}{\downarrow}

\begin{document}

\title{Mott-Anderson Metal-Insulator Transitions from Entanglement}

\author{G. A. Canella}
\affiliation{Institute of Chemistry, S\~{a}o Paulo State University, 14800-090, Araraquara, S\~{a}o Paulo, Brazil}
\author{V. V. Fran\c{c}a}

\affiliation{Institute of Chemistry, S\~{a}o Paulo State University, 14800-090, Araraquara, S\~{a}o Paulo, Brazil}

\begin{abstract} 

A metal can be driven to an insulating phase through distinct mechanisms. A possible way is via the Coulomb interaction, which then defines the Mott metal-insulator transition (MIT). Another possibility is the MIT driven by disorder, the so-called Anderson MIT. Here we analyze interacting particles in disordered Hubbard chains $-$ thus comprising the Mott-Anderson physics $-$ by investigating the ground-state entanglement with density functional theory. The localization signature on entanglement is found to be {a decreasing on the entanglement degree in comparison to the clean (without interaction and disorder) case, with} local minima at certain critical densities. Individually, the Mott (Anderson) MIT has a single critical density whose minimum entanglement decreases as the interaction (disorder) enhances. While in the Mott MIT entanglement saturates at finite values, characterizing partial localization, in the Anderson MIT the system reaches full localization, with zero entanglement, for sufficiently strong disorder. In the combined Mott-Anderson MIT, we find three critical densities referring to local minima on entanglement. One of them is the same as for the Anderson MIT, but now the presence of interaction requires a stronger disorder potential to induce {full} localization. A second critical density is related to the Mott MIT, but due to disorder it is displaced by a factor proportional to the concentration of impurities. The third local minimum on entanglement is unique to the concomitant presence of disorder and interaction, found to be related to an effective density phenomenon, thus referred to as a Mott-like MIT. Since entanglement has been intrinsically connected to the magnetic susceptibility $-$ a quantity promptly available in cold atoms experiments $-$ our detailed numerical description might be useful for the experimental investigation of Mott-Anderson MIT.

\end{abstract}


\maketitle

\section{Introduction}

A metal can be transformed into an insulating system via different mechanisms. As proposed by Mott \cite{16ref1}, the long-range character of the Coulomb interaction may lead to the metal-insulator transition (MIT). Also the short-range electron-electron interaction may induce MIT when there is one electron per lattice site, as proposed by Hubbard \cite{17ref1}. These transitions driven by interactions \cite{23, 25, 27} are commonly referred to as Mott MIT. In the absence of interactions, the metal-insulator transition may alternatively be induced by disorder \cite{14ref1}, the so-called Anderson MIT, due to coherent backscattering from randomly distributed impurities. 

Theoretically, despite the difficulty to approach the MIT in solid systems, considerable progress has been achieved in cases where both disorder and electronic correlations play an important role \cite{1, 2, 5, 7, 8, 10, 21, 22}. However the interplay Mott-Anderson MIT is far from being completely understood. Besides, most of the studies applies dynamical mean-field theory (DMFT) \cite{1ref6}, which allows the treatment of correlations and randomness on the same footing but is computationally demanding and restricted to rather simple systems.

A more general approach consists on exploring the differences between metal and insulator through the distribution of electrons in their many-body ground state, as qualitatively proposed by Kohn \cite{6ref6, 7ref6} and more recently applied to the MIT context \cite{8ref6, 9ref6, 10ref6, 11ref6, 6}.  This then suggests that the electronic density function and other related density functional quantities, within density functional theory (DFT) \cite{dft}, could be useful for describing MIT. Although it is well known that first-principles investigations via standard DFT methodology encounter considerable difficulty to recover the many-body gap, there are several proposals to correct this problem and improve the MIT description via DFT \cite{30, 31, fvc, ent3,11, 12, 13, 15}.  

Moreover, recently DFT has been applied in the context of disordered superfluids to detect the superfluid-insulator transition (SIT) \cite{gui1, gui2}. In this case the transition was probed via entanglement, widely recognized as a powerful tool for identifying quantum phases transitions \cite{qpt1, qpt2,qpt3,qpt4, qpt5}, including exotic states of matter \cite{exo1, exo2, exo3, exo4, exo5}. Additionally, entanglement has been used to explore Mott MIT, without disorder \cite{18, 19, ent4, tobi}.  Entanglement was also found to be intrinsically connected to the magnetic susceptibility \cite{chi1, chi2, chi3}, which is promptly available in cold atoms experiments and thus it could allow  the experimental investigation of Mott-Anderson MIT.

In spite of that only a few studies uses entanglement to investigate Mott-Anderson MIT \cite{ent1, ent2, francaAmico2011}. Refs. \cite{ent1} and \cite{ent2} adopt density matrix renormalization group (DMRG) techniques, numerically exact, but a highly costly method, in particular for disordered systems which require a large number of random disorder configurations. Ref. \cite{francaAmico2011} uses instead an approximate DFT approach, allowing faster calculations, but the MIT was barely investigated, it was actually used as a test bed for the proposed  analytical density functional for entanglement. Thus a deeper investigation of systems with Mott and Anderson physics via entanglement is still missing.

Here we apply DFT calculations to obtain entanglement and investigate the Mott-Anderson MIT in disordered Hubbard chains. We find that entanglement {decreases with respect to the clean (no disorder, no interaction) case, with a single} local minimum at a certain critical density {when} the Mott and the Anderson MIT {are considered} individually. While the Mott-MIT entanglement minimum saturates at a finite value, characterizing a partially localized state, the Anderson-MIT minimum reaches zero for sufficiently strong disorder, characterizing then full localization. In the presence of both disorder and interaction, we find {\it i)} that the Mott critical density is displaced by a factor proportional to the concentration of impurities, {\it ii)} the Anderson critical density is maintained, but the presence of interaction requires a stronger disorder potential to induce {full} localization, and {\it iii)} there is a third critical density appearing exclusively in the presence of both, disorder and interaction, which is related to a Mott-like MIT due to an effective density phenomenon. 


\section{Methods}

We consider interacting fermions {at zero temperature} as described by the one-dimensional Hubbard model with on-site disorder, 

\be{}
H = -t\sum_{\la ij \ra \sigma}(\hat{c}^{\dagger}_{i\sigma}\hat{c}_{j\sigma}+ h.c.) + 
U\sum_i\hat{n}_{i\ua}\hat{n}_{i\da} + \sum_{i\sigma}V_i\hat{n}_{i\sigma},
\ee{}
where $\hat{c}^{\dagger}_{i\sigma}$ ($\hat{c}_{i\sigma}$) is the creation (annihilation) operator and $\hat{n}_{i\sigma} = \hat{c}^{\dagger}_{i\sigma}\hat{c}_{i\sigma}$ is the particle density operator, with $z$-spin component $\sigma = \ua,\da$ at site $i$. The average density or filling factor is $n = N/L$, where $N = N_{\ua} + N_{\da}$ is the total number of particles and $L$ is the chain size. We adopt spin-balanced populations, $N_{\ua} = N_{\da}$, $L = 100$, open boundary conditions and $U>0$. We express the local Coulomb interaction $U$ and the disorder potential $V_i$ in units of the hopping parameter $t$, and set $t=1$.

\begin{figure} [!t]
    \centering
        \includegraphics[scale=.31]{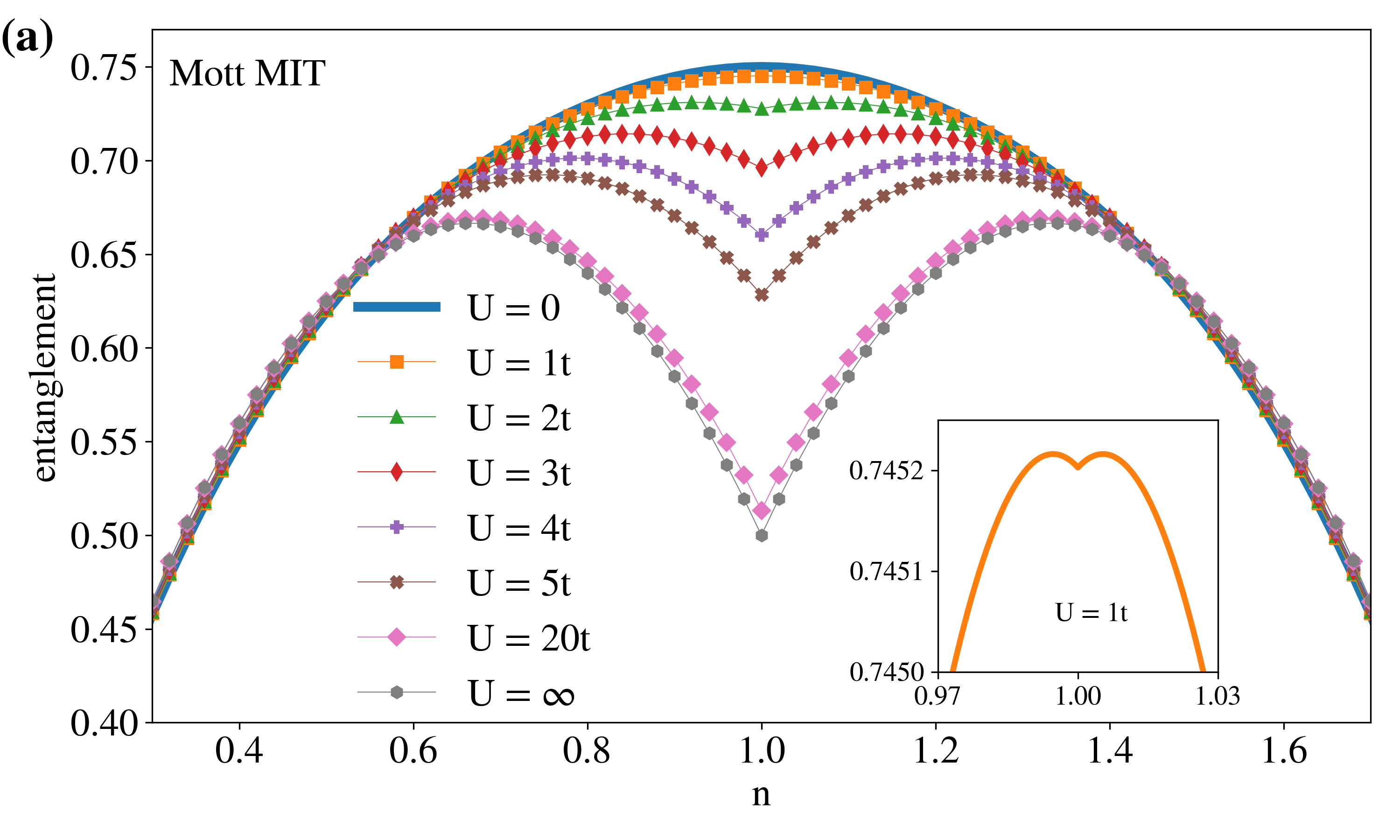}\\
         \includegraphics[scale=.31]{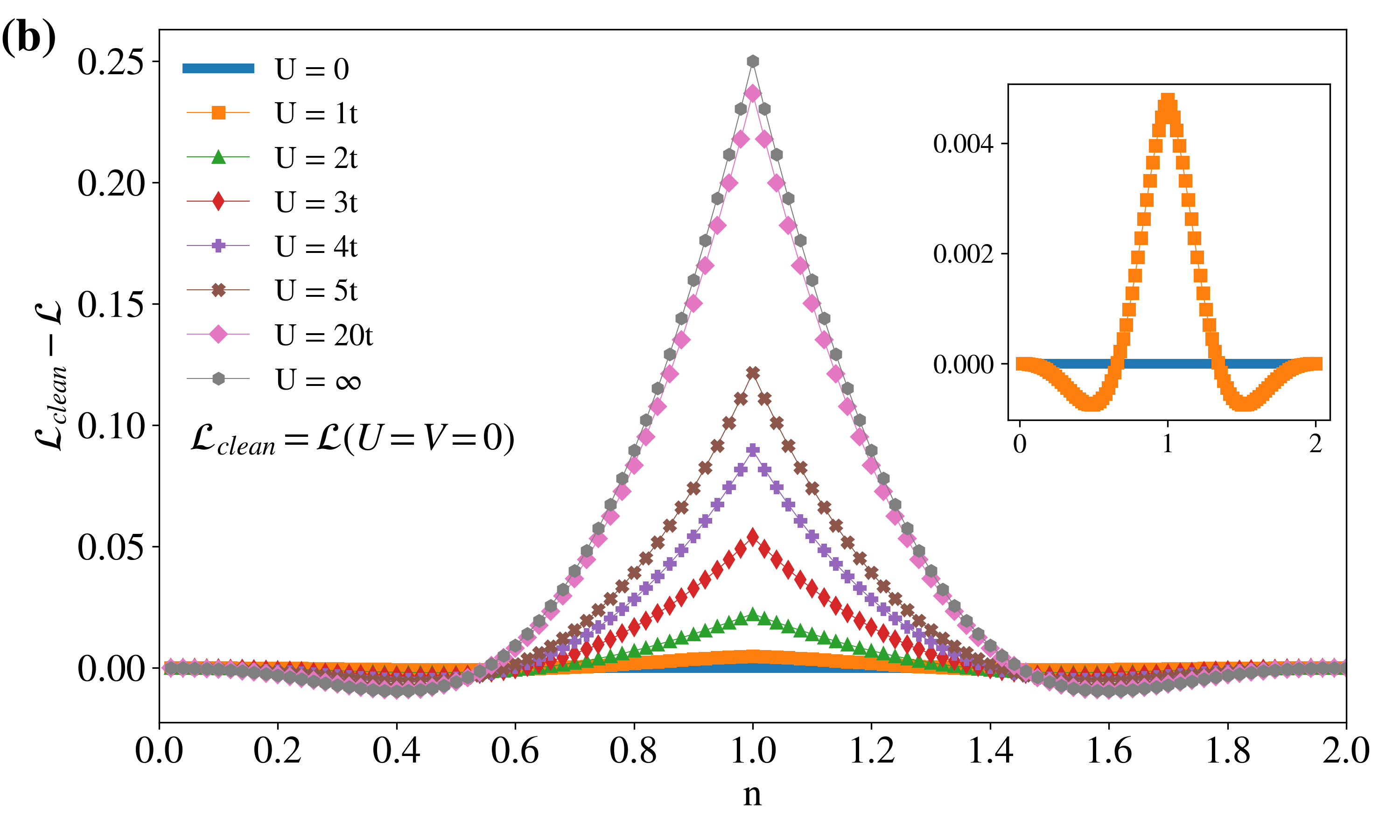}\\
  
    \caption{{(a) Entanglement $\mathcal L$ as a function of the density for interacting non-disordered nanostructures (Mott MIT). (b) Entanglement difference  $\mathcal L_{clean} - \mathcal L$, where $\mathcal L_{clean}\equiv \mathcal L(n,U=0,V=0)$. Insets for $U=t$.} } 
\end{figure}

We consider pointlike impurities of strength $V$ randomly distributed along the chain within a certain concentration $C\%$, defined as $C \equiv 100L_V/L$ where $L_V$ is the number of impurity sites. To ensure results that are independent on specific configurations of impurities, we generate $100$ samples for each set ($C,V;U,n$) of parameters, and thus all quantities analyzed here are an average over these 100 random disorder configurations. This huge amount of data makes it impracticable to use exact methods such as DMRG, so we apply instead DFT calculations for the Hubbard model {(see \cite{dft1-a,dft1-b,dft1-d} for a review on the accuracy of this formalism)}, obtaining the per-site ground-state energy $e_0$ and the density profile. 

We explore the Mott-Anderson MIT via the average single-site entanglement, which is the entanglement between each site with respect to the remaining $L-1$ averaged over the sites. The ground-state entanglement of homogeneous ($V=0$, $L=\infty$) chains can be quantified by the linear entropy, 
\begin{equation}
\mathcal L^{hom}=1-\text{w}_{\uparrow}^2-\text{w}_{\downarrow}^2-\text{w}_{2}^2-\text{w}_{0}^2,
\end{equation}
where $\text{w}_{\uparrow}=\text{w}_{\downarrow}=n/2-{\text{w}_2}$ are unpaired probabilities, $\text{w}_{0}=1-\text{w}_{\uparrow}-\text{w}_{\downarrow}-\text{w}_2$  {is the empty occupation probability and $\text{w}_{2}=\partial e_0/\partial U$ the paired probability, the latter quantified through a parametrization \cite{lsoc} for the Lieb-Wu exact energy \cite{lw}.} The maximum entanglement, $\mathcal L=0.75$, occurs when the four occupation states are equally probable: $\text{w}_{\uparrow}=\text{w}_{\downarrow}=\text{w}_{2}=\text{w}_{0}=1/4$. 

For the inhomogeneous (disordered and finite) chains, we {adopt instead an} approximate density functional for the linear entropy $\mathcal L^{hom}(n,U>0)$ \cite{francaAmico2011} to obtain entanglement via DFT within a local density approximation \cite{vivi}
\begin{equation}
\mathcal L\approx \mathcal L^{DFT}\equiv \frac{1}{L}\sum_i\mathcal L^{hom}(n,U>0)|_{n\rightarrow n_i}.
\end{equation}
This approach \cite{sm} has also been successfully used recently to explore the superfluid-insulator transition in disordered superfluids \cite{gui1, gui2}.

\section{Results and Discussion}

\begin{figure}[t!]
    \centering
        \includegraphics[scale=.31]{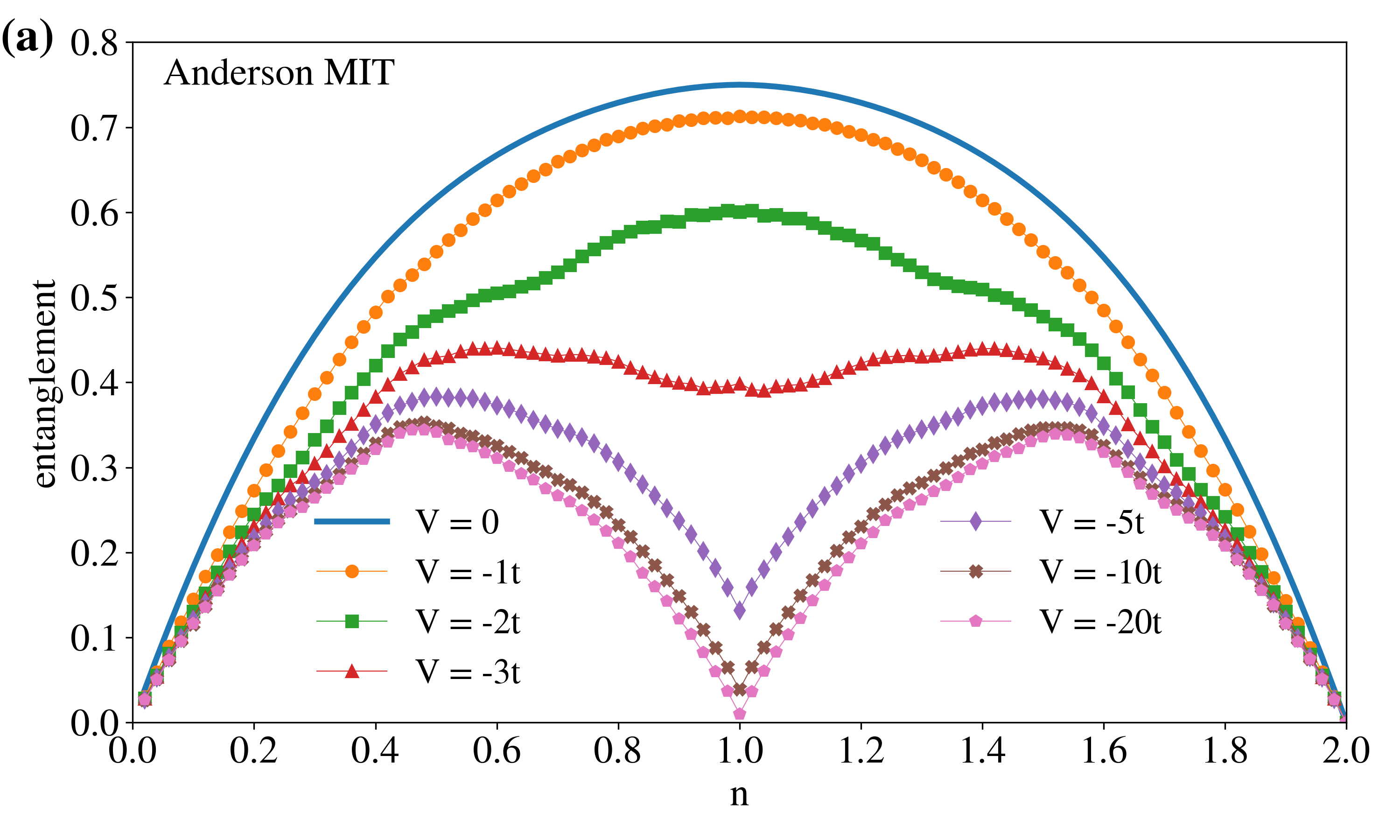}\\
             \includegraphics[scale=.31]{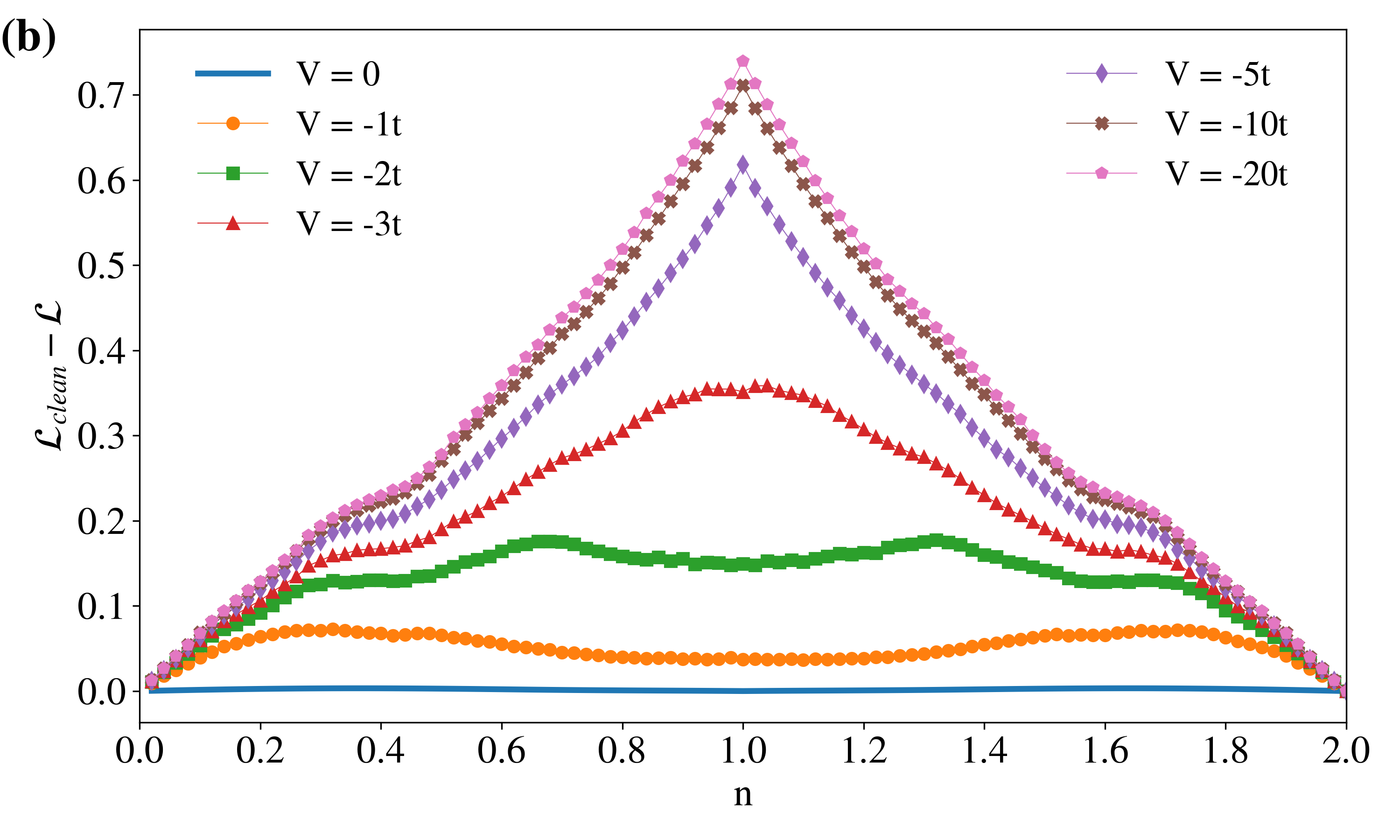}\\
    \caption{{(a) Entanglement $\mathcal L$ as a function of the density for disordered non-interacting nanostructures (Anderson MIT). (b) Entanglement difference  $\mathcal L_{clean} - \mathcal L$, where $\mathcal L_{clean}\equiv \mathcal L(n,U=0,V=0)$. In all curves $C=50\%$.}}
\end{figure}

We start by analyzing the behavior of entanglement with respect to the average density in the two clean cases: the Mott MIT for $V=0$ and the Anderson MIT for $U=0$. In Figure 1{a}, for the Mott MIT, we find that as the interaction increases, $\mathcal L$ decreases near $n=1$. In particular, {for all $U>0$} a local minimum at the critical density $n_C^U=1$ emerges and becomes more pronounced when $U$ increases, with entanglement saturating at the finite value $\mathcal L=0.5$ for $U\rightarrow\infty$. This minimum entanglement at $n_C^U=1$ is the signature of the Mott MIT \cite{us}: the interaction freezes the translational degrees of freedom with maximum single-occupation probabilities ($\text{w}_{\uparrow}=\text{w}_{\downarrow}\rightarrow 0.5$) and no further effect by enhancing $U$. This remaining entanglement, $\mathcal L (U\rightarrow \infty)=0.5$, is associated to the spin degrees of freedom, which are not frozen out, so in the Mott MIT the system reaches a partially localized state. {By quantifying the entanglement decreasing with respect to the clean case via the difference $\mathcal L_{clean}-\mathcal L$, in Figure 1b, one confirms that the decreasing is essentially concentrated near $n_C^U=1$.}

{In contrast, in Figure 2a} for the Anderson MIT, we find that entanglement decreases with disorder {for all the densities. This overall entanglement decreasing characterizes the Anderson localization: the system is localized for $V>0$, what is also confirmed in Figure 2b by the fact that the difference  $\mathcal L_{clean}-\mathcal L>0$  for any $V, n$. Fig. 2a also reveals a minimum entanglement (corresponding to a maximum $\mathcal L_{clean}-\mathcal L$ in Fig. 2b) at} the critical density $n_C^V=2C/100$ for $|V|\gtrsim 3t$.  This critical $n_C^V$, first observed at the superfluid-insulator transition \cite{gui1, gui2}, corresponds to the case where the number of impurity sites for $V<0$ (non-impurity sites for $V>0$) is equal to twice the number of particles. Thus the entanglement drop at $n_C^V$ characterizes the {full localization in the} Anderson MIT: the attractive disorder freezes the degrees of freedom by favoring the double occupancy at the impurity sites (for $V>0$ the non-impurity sites are doubly occupied at $n_C^V=1-2C/100$). {Notice that} differently from the Mott case, in the Anderson MIT the system may reach a fully localized state if disorder is sufficiently strong to make $\text{w}_{2}\rightarrow 1${, such that} $\mathcal L(|V|\rightarrow \infty)=0$.

\begin{figure}
    \centering
        \includegraphics[scale=.31]{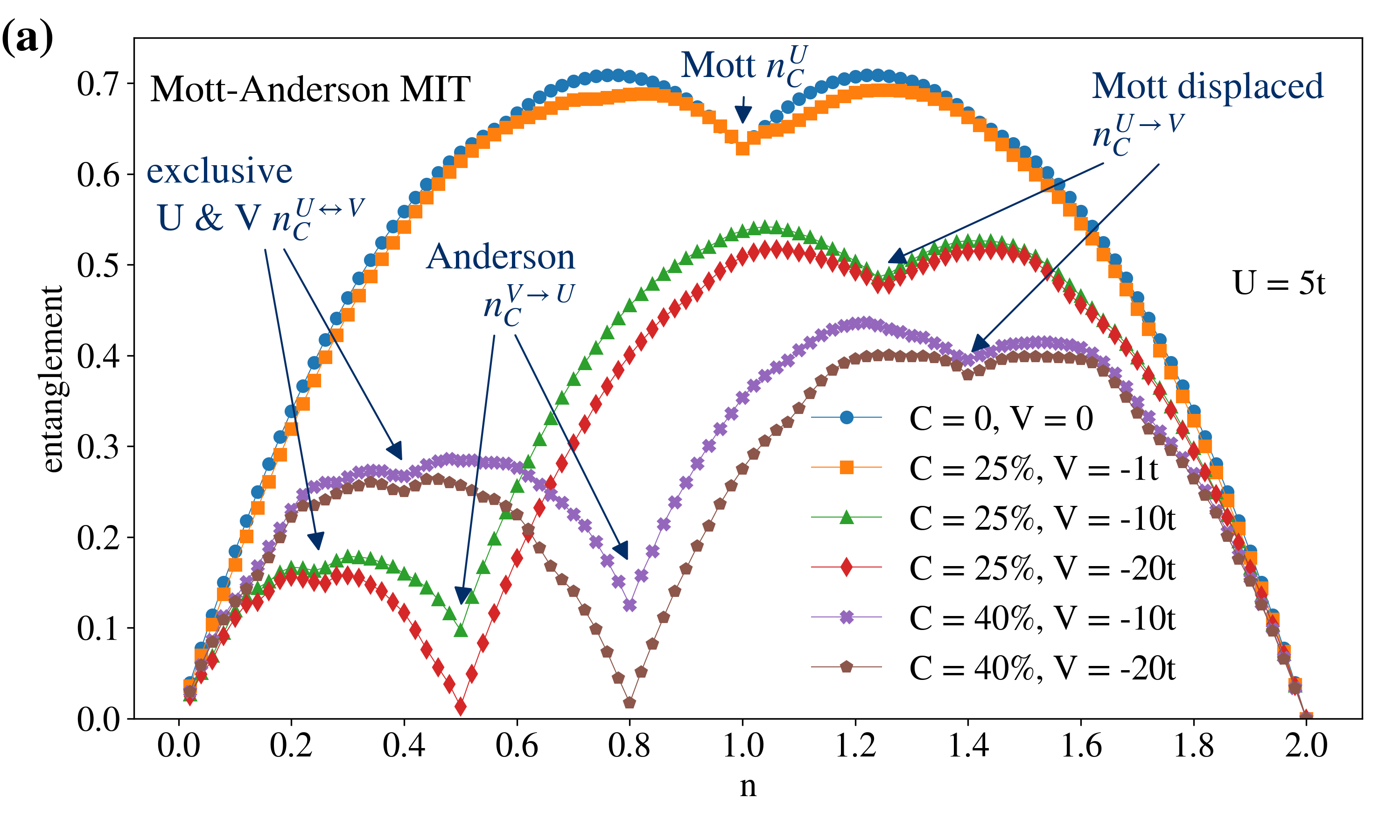}\\
            \includegraphics[scale=.31]{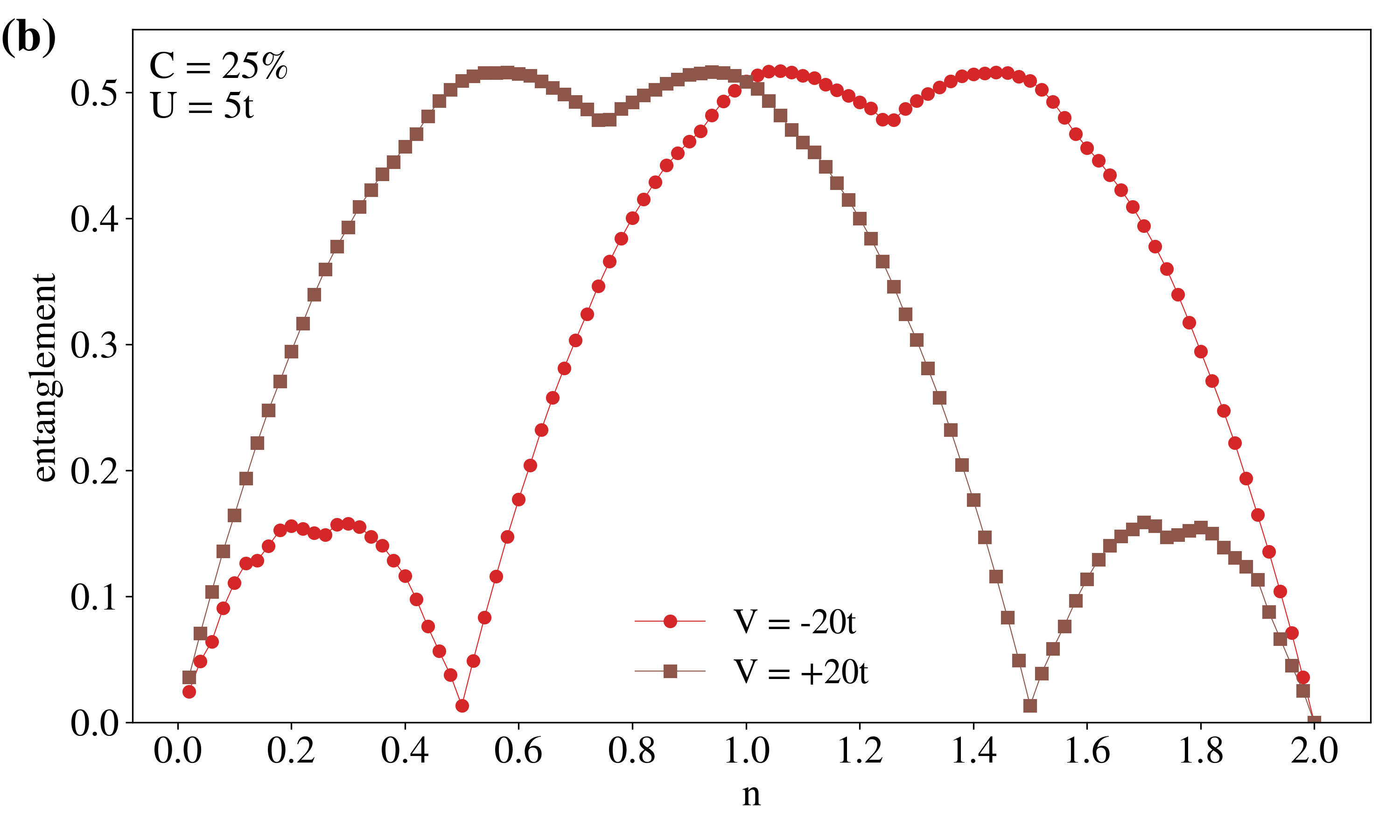}
    \caption{{(a) Entanglement $\mathcal L$ as a function of the density for disordered interacting nanostructures (Mott-Anderson MIT).} The critical density {$n_C^V=2C/100$ for the clean Anderson MIT case} is maintained in the presence of $U$: $n_C^{V\rightarrow U}=n_C^V$, albeit with higher entanglement $\mathcal L$. The critical density for the clean Mott MIT $n_C^U=1$ in the presence of {sufficiently strong} $V<0$ is displaced to $n_C^{U\rightarrow V}=1+C/100$. An extra critical density appears exclusively in the presence of both $U$ and $V$, $n_C^{U\leftrightarrow V}=C/100$, related to a Mott-like MIT. {(b) Particle-hole symmetry when changing from $-V$ to $+V$: the critical densities can be obtained by replacing $n_C\rightarrow \tilde n_C= 2-n_C$.}  }  
\end{figure}

In Figure 3{a} we then turn on both disorder and interaction and investigate the impact of $U$ on the Anderson MIT and of $V$ on the Mott MIT.  {We find} that for small disorder strength ($V=-1t$) there is almost no influence on the Mott MIT: the minimum entanglement is maintained at $n_C^U=1$ and the {full localization minimum} at $n_C^V$ in the Anderson MIT does not appear, as in the clean case. As $V$ increases, we observe {the impacts} of the disorder on the Mott MIT:  {{\it i)} the entanglement at $n_C^{U\rightarrow V}$ is not limited to $\mathcal L>0.5$ (as in the clean interacting case), since now $V$ can also freezes the spin degrees of freedom, and {\it ii)}} the critical density for the Mott transition is displaced to $n_C^{U\rightarrow V}>1$. This shift is reflecting the fact that the effective density in the metallic non-impurity sites, $n_{V=0}$, is smaller than $n$, since the particles are attracted to the $V<0$ sites, such that it is necessary $n>1$ to induce the Mott MIT at $n_{V=0}=1$. To determine $n_C^{U\rightarrow V}$ we start by considering that for $V<0$ most of the particles accumulates in the attractive $L_V$ impurity sites (up to 2 per site, so $2L_V$ particles), thus leaving an effective $N-2L_V$ number of particles in the remaining $L-L_V$ metallic sites. So the effective density can be defined as  $n_{V=0}\equiv (N-2L_V)/(L-L_V)$ and the critical $n_C^{U\rightarrow V}$ will be then the average density $n$ for which $n_{V=0}=1$. In terms of the concentration {this} can be written as {$n_C^{U\rightarrow V}=1+C/100$,} which agrees with the numerical results of Fig. 3{a}. 

\begin{table}[!t]
\caption{Critical densities for {local entanglement minima} for $V<0$ in the clean cases, $n_C^U$ for Mott and $n_C^V$ for Anderson, and in the combined Mott-Anderson MIT: $n_C^{U\rightarrow V}$ for the impact of $V$ in the $n_C^U$,  $n_C^{V\rightarrow U}$ for the impact of $U$ in the $n_C^V$, and $n_C^{U\leftrightarrow V}$ for the Mott-like MIT that only appears in the presence of $V$ and $U$. For $V>0$ the corresponding critical densities can be obtained via a particle-hole transformation, replacing the desired critical density $n_C\rightarrow \tilde n_C=2-n_C$.}
\vspace{0.3cm}
\begin{tabular}{p{3.5cm}p{4cm}}

\hline
&\\[-6pt]
 Mott MIT & $n_C^U=1$\\
\small{$U>0$}, $V=0$ & \\[3pt]
\hline
&\\[-6pt]
Anderson MIT &$n_C^V=2C/100$\\
 \small $|V|\gtrsim 3t$, $U=0$ & \\[3pt]
 
\hline
&\\[-6pt]
Mott-Anderson MIT & $n_{C}^{U\rightarrow V}=1+C/100$ \\[3pt]
\small {$U>0$}, $|V|\gtrsim 3t$&   $n_C^{V\rightarrow U}=n_C^V=2C/100$ \\[3pt]
 &  $n_C^{U\leftrightarrow V}=C/100$\\
&\\[-6pt]

\hline
\end{tabular}
\end{table}

Notice however that a similar effective density phenomenon occurs also for the impurity sites: for sufficiently strong $V<0$, there will be an average density $n<1$ for which the impurity-sites effective density, $n_{V}\equiv N/L_V$, will be equal to 1, since $n=N/L<n_{V}$. This then means that there is another critical density {$n_C^{U\leftrightarrow V}=C/100$} associated to a {\it Mott-like MIT} induced by the disorder. Indeed we see in all curves of Fig. 3{a} a smoother minimum at $n_C^{U\leftrightarrow V}$ which was absent in the clean cases and is related to the Mott-like MIT induced by sufficiently strong disorder. Other effective-density features  {have} been also reported for Hubbard chains in the presence of binary-alloy disorder \cite{10}. 

Concerning the impact of the interaction on the Anderson MIT, Fig. 3a reveals that albeit the critical density remains at $n_C^V$, i. e. $n_C^{V\rightarrow U}=n_C^V=2C/100$, the minimum entanglement is higher (compare it to the curve $V=-10t$ in Fig. 2{a}). This then shows that in the presence of $U$ it is required a stronger disorder potential for {the full localization}, marked by $\mathcal L=0$. This comes from the fact that $U$ and $V$ contribute differently to the insulating phase: while $U>0$ contributes by favoring the unpaired probabilities ($\text{w}_{\uparrow}$, $\text{w}_{\downarrow}$), $V$ contributes by favoring the doubly-occupied probability ($\text{w}_{2}$) \cite{foot1}. Thus the competition between $U$ and $V$ requires a stronger disorder potential for the {full localization in the} Anderson MIT. Accordingly, for the superfluid-insulator transition driven by disorder, where the attractive interaction $U<0$ also favors the double occupancy, this competition effect is not observed \cite{gui1, gui2}.

Fig. 3a also shows that for a fixed $V$, the minimum $\mathcal L$ at $n_C^{V\rightarrow U}$ becomes smaller for lower concentrations. This occurs because the critical density is smaller for lower $C$ and the impact of the interaction $U$ becomes less relevant for low densities, thus diminishing the $U$ and $V$ competition effect. {Additionally,} for $n<n_C^{V\rightarrow U}$ entanglement increases but reaches a lower $\mathcal L$ platform in comparison to entanglement for $n>n_C^{V\rightarrow U}$. This phenomenon is related to the fact that for small densities the disorder hampers the connections among the particles, decreasing $\mathcal L$. Consistently, as $C$ enhances, the average distance among the impurity sites (for $V<0$) decreases, and the platform becomes higher and broader. 

Finally, we emphasize that all the MIT critical densities obtained here for attractive disorder, summarized in Table I, can be obtained for repulsive disorder by replacing the desired critical density $n_C\rightarrow \tilde n_C= 2-n_C$ due to the particle-hole symmetry when changing from $-V$ to $+V$. This is confirmed by Fig. 3{b}, which shows the three entanglement minima {for the Mott-Anderson with $V>0$} at the replaced critical densities $\tilde n_C^{V\rightarrow U}=2-2C/100$ {(Anderson)}, $\tilde n_C^{U\rightarrow V}= 1-C/100$ {(Mott)} and $ \tilde{n}_C^{U\leftrightarrow V}=2-C/100$ {(Mott-like).}

\section{Conclusions}

In summary, we have investigated the Mott-Anderson MIT via entanglement of interacting particles in disordered Hubbard chains. We find that {localization is marked by an entanglement decreasing, with local minima at the critical densities summarized in Table I.} While separately the Mott and Anderson MIT have only a single critical density each, for the combined Mott-Anderson MIT we find three distinct critical densities. One of them is the clean Anderson-MIT critical density, but the presence of interaction now requires a stronger disorder strength to localize the system with the same entanglement minimum.  Another one is the clean Mott-MIT critical density but displaced by a factor proportional to the concentration of impurities. The third is a critical density exclusive from the interplay between disorder and interaction, and is found to be related to an effective density phenomenon, thus referred to as a Mott-like MIT.  

\begin{center}\bf ACKNOWLEDGMENTS \end{center}

{We thank M.C.O. Aguiar for fruitful discussions.} VVF was supported by FAPESP (2019/15560-8) {and GAC by} the Coordena\c{c}\~{a}o de Aperfei\c{c}oamento de Pessoal de Nivel Superior - Brasil (CAPES) -
Finance Code 001.



\begin{thebibliography}{99}
\bibitem{foot1} This is the case for both, attractive and repulsive disorder: for $V<0$ $\text{w}_2$ is favored in the impurity sites, while for $V>0$ it is favored in the non-impurity sites.
\end{thebibliography}
\end{document}